\documentclass[prl,reprin,amsmath,amssymb,showpacs,twocolumn]{revtex4-1}

\usepackage[dvipdfm]{graphicx}
\usepackage{bm}

\usepackage{mathrsfs}
\usepackage{color}

\DeclareMathOperator{\diag}{diag}

\begin{document}
\title{Topological Blount's theorem of odd-parity superconductors}
\author{Shingo Kobayashi$^1$}
\author{Ken Shiozaki$^2$}
\author{Yukio Tanaka$^1$}
\author{Masatoshi Sato$^1$}
\affiliation{$^1$Department of Applied Physics, Nagoya University, Nagoya 464-8603, Japan \\ $^2$Department of Physics, Kyoto University, Kyoto 606-8502, Japan}

\date{\today}

\begin{abstract}
   From the group theoretical ground, the Blount's theorem prohibits the
 existence of line nodes for odd-parity superconductors (SCs) in the
 presence of spin-orbit coupling (SOC). We study the topological
 stability of line nodes under inversion symmetry.
From
 the topological point of view, we renovate the stability condition of
 line nodes, in which we not only generalize the original statement,
 but also establish the relation to zero-energy surface flat
 dispersions. The topological instability of line nodes in odd-parity SCs
 implies not the absence of bulk line nodes but
 the disappearance of the corresponding 
 zero-energy surface flat dispersions due to surface Rashba SOC,
 which gives an experimental means to distinguish line nodes in
 odd-parity SCs from those in other SCs. 
\end{abstract}
\pacs{}
\maketitle
{\it Introduction.}---
Nontrivial nodal structures are salient feature of unconventional
SCs. The existence of nodes gives a clue to symmetry of Cooper
pairings and has an influence on power low behaviors of the temperature
dependences such as the specific heat, the NMR relaxation rates, and so
on \cite{Legget:1975,Sigrist:1991}. In 1980s, heavy fermion materials
have attracted much attention as a candidate of unconventional SCs, and
then the Cooper pairs were classified based on the group theory because
ones break not only gauge symmetry but also crystal symmetries due to
SOC and crystal
fields~\cite{Anderson:1984,Sigrist:1991,Blount:1985,Volovik:1985,Ueda:1985}. By
this means, Blount proved the impossibility of line nodes in odd-parity
SCs \cite{Blount:1985}. Assuming a time-reversal-invariant
single-band spin-triplet Cooper pair,
he showed that a large region of zero gap is ``vanishingly
improbable'' in the presence of SOC.
The statement is now known as Blount's
theorem. To the contrary, however, real candidate materials of heavy
fermion odd-parity SCs such as UPt$_3$~\cite{Joynt:2002} often have suggested
the existence of line nodes experimentally.
This is because the influence of SOC on bulk Cooper pairs is strongly suppressed
by the Fermi energy \cite{Yanase:2003}. 
Hence, the validity of the Blount's theorem has been suspicious. 

While the group theoretical arguments seem not to work well, 
there is another arguments for the stability of line
nodes~\cite{Sato:2006, Beri:2010},
in which the nodal structures are classified by
topological
invariants~\cite{Volovik:2003,Horava:2005,Zhao:2013v1}. 
Without
assuming a large SOC, this method enables us to treat both symmetric and
accidental nodes in a unified way and moreover includes the influence of
normal states and multiband structures. Therefore, the topological
classification has a potential to extend the original Blount's
theorem and to fill the gap between the group theoretical classification
and the experimental measurement.
In addition, topologically stable line nodes can manifest themself zero
energy surface flat dispersions via the bulk-boundary correspondence at
certain surfaces~\cite{Sato:2011,Yada:2011,Tanaka:2012,Schnyder:2011,Brydon:2011,Schnyder:2012,Schnyder:2013,Matsuura:2013}.

\begin{figure}[tbp]
\centering
 \includegraphics[width=8cm]{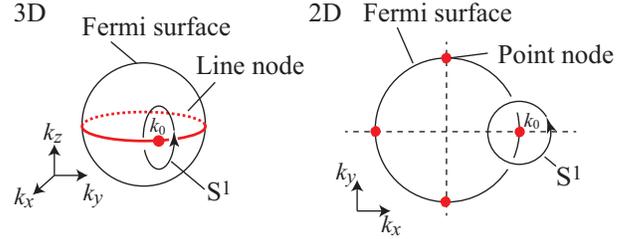}
 \caption{(color online) The line node of the $p_z$-wave SC in 3D (left) and the point node of the $d_{xy}$-wave SC in 2D (right). $\bm{k}_0$ indicates the position of the node. In both cases, $S^1$ wraps around the node.}\label{fig:line-node}
\end{figure}

In this letter, we present a topological version of the Blount's theorem in
terms of 
the topological K-theory and the Clifford module
\cite{Atiyah:1964,Karoubi:1978}, and discuss the stability of nodes by
utilizing the Clifford algebra extension
method~\cite{Kitaev:2009,Stone:2011,Wen:2012,Morimoto:2013,Abramovici:2012}. 
The purpose of this letter is to prove the following statements: (i){\it
A line node in odd-parity SCs is
topologically unstable with or without TRS in the absence of additional
symmetry.}
(ii) {\it An additional
symmetry such as mirror symmetry or spin-rotation symmetry (SRS) may stabilize
a bulk line node in odd-parity SCs, but the corresponding surface flat
dispersion is fragile and disappears due to surface Rashba SOC.}  
Here the original Blount's theorem is included in the statement (i).
In the heavy fermion materials, the statements (i) and (ii) imply
that zero energy surface flat dispersions in odd-parity SCs with line
nodes break down easily.
Hence, it is
possible to distinguish odd-parity Cooper pairs from even-parity ones
or noncentrosymmetric ones
by the behavior of the zero-energy flat dispersion. 
\begin{table}[tbp]
\centering
 \caption{Classification of nodes, which occurs at $\bm{k}_0 \neq -\bm{k}_0 + \bm{G}$, in the system with $C$, $T$, and $\tilde{P}$. The first, second, third, fourth columns show the symmetry classes, the symmetry constraint for each class, the parity of gap functions, and the classifying space $\mathcal{Q}$, respectively. The following columns show the topological classification for $p=0$, $1$, and $2$. In 3D, each codimension represents a surface node, a line node, and a point node, respectively.} \label{tab:parity}
 \begin{tabular}{ccccccc} 
\hline \hline
Class & Symmetry & Parity  &  $\mathcal{Q}$   & $p=0$  & $p=1$ & $p=2$  \\ \hline
D      &  \{1\}    & N/A   &  $C_0$ &  $\mathbb{Z}$ & $0$ & $\mathbb{Z}$ \\
DIII    & \{$CT$\}&  N/A  &   $C_1$ & $0$  &  $\mathbb{Z}$ & $0$   \\ \hline
P+ D & \{$C\tilde{P}$\} & even  &  $ R_2 $ & $\mathbb{Z}_2$ & $0$ & $\mathbb{Z}$ \\
       &  &  odd & $ R_6$ & $0$ & $0$ & $\mathbb{Z}$  \\ \hline
P+ DIII & \{$C\tilde{P},T\tilde{P}$\}& even & $R_3 $ &  $0$ & $\mathbb{Z}$ & $0$  \\ 
                &  & odd & $R_5 $ & $0$ & $0$ & $0$  \\ 
 \hline \hline
 \end{tabular}
\end{table}    
  
{\it Formulation.}---
We start with the Bogoliubov de Genne (BdG) Hamiltonian:
\begin{align}
 H = \frac{1}{2} \sum_{\bm{k},\alpha, \alpha'} \left( c_{\bm{k} \alpha}^{\dagger}, c_{-\bm{k} \alpha} \right) H(\bm{k}) \left( \begin{array}{@{\,} c @{\,}} c_{\bm{k} \alpha'} \\ c_{-\bm{k} \alpha'}^{\dagger} \end{array}\right) , \label{eq:BdG1}
\end{align}
where $H(\bm{k})$ is given by
\begin{align}
 H(\bm{k}) = \begin{pmatrix} \epsilon (\bm{k})_{\alpha \alpha'} & \Delta (\bm{k})_{\alpha \alpha'} \\ \Delta (\bm{k})_{\alpha \alpha'}^{\dagger} & -\epsilon (-\bm{k})_{\alpha \alpha'}^{T} \end{pmatrix}. \label{eq:BdG2}
\end{align}
$c_{\bm{k}\alpha}^{\dagger}$ ($c_{\bm{k}\alpha'}$) represents the
creation (annihilation) operator of electron with momentum $\bm{k}$. The
suffix $\alpha$ labels other degrees of freedom such as spin, orbital,
sublattice indices, and so on. $\epsilon (\bm{k})_{\alpha \alpha'}$ and
$\Delta (\bm{k})_{\alpha \alpha'}$ are the Hamiltonian in the normal state
and the gap function, respectively. In the case of a single-band spin-triplet
Cooper pair, the gap function is given by $\Delta (\bm{k}) = i
\bm{d}(\bm{k}) \cdot \bm{\sigma} \sigma_y $, where $\bm{\sigma}$ is the
Pauli matrix and the $\bm{d}$ vector satisfies $\bm{d}(-\bm{k})=
-\bm{d}(\bm{k})$ ($d_i \in \mathbb{R}$ ($i=x,y,z$) if TRS exists). The
BdG Hamiltonian naturally has particle-hole symmetry (PHS) such that
\begin{align}
 CH(\bm{k})C^{\dagger} = -H(-\bm{k}), \ \ C^2 = 1. \label{eq:PHS}
\end{align}
Also, TRS is defined by
\begin{align}
T H(\bm{k}) T^{\dagger} = H(-\bm{k}), \ \ T^2 = -1, \label{eq:TRS}
\end{align}
where $C$ and $T$ are antiunitary. In addition, we assume inversion
symmetry (IS) such that
\begin{align}
 &P \epsilon (\bm{k}) P^{\dagger} = \epsilon (-\bm{k}), \ \ P \Delta (\bm{k}) P^T = \eta^P_C \Delta (-\bm{k}), \label{eq:IS}
\end{align}
where $P$ acts on the creation (annihilation) operator as
$c_{\bm{k}\alpha}^{\dagger}$ $\to$ $P_{\alpha \alpha' }^{\ast}
c_{-\bm{k}\alpha'}^{\dagger}$ ($c_{\bm{k}\alpha'}$ $\to$ $P_{\alpha
\alpha'} c_{-\bm{k}\alpha'}$) and satisfies $P^2 = 1$. The factor
$\eta^P_C $ specifies either even-parity ($\eta^P_C =
1$) or odd-parity ($\eta^P_C = -1$) of the gap function. In the Nambu
representation, we denote $P$ as $ \tilde{P} = \diag ( P , \eta^P_C
P^{\ast}) $ \cite{Sato:2010}. 
The parity of the gap function determines the commutation or
anti-commutation relation between $C$ and $\tilde{P}$: $[C,
\tilde{P}]=0$ ($\{C, \tilde{P}\}=0$) for even-parity (odd-parity) pairings.  
Note also that $[T,\tilde{P}] =0$ since $P$ does not act on the spin space.

{\it Stability of node and symmetry.}---
A node of SCs is a set of ${\bm k}$ satisfying $\det H(\bm{k}) =
0$.
In $d$-dimensions, the node with codimension $p+1$ defines a
$(d-p-1)$-dimensional submanifold $\Sigma$.
For example, a line node in three-dimensions has codimension 2, and it
defines one-dimensional manifold along the node.
%
If we consider a symmetry preserving small
perturbation of $H$,
the node either shifts its position slightly or goes away
completely due to the emergence of a gap. The former implies that the
node is topologically stable since we cannot remove it by any small
perturbations.

To define the topological stability of the node precisely, consider a small $p$-dimensional sphere $S^p$
wrapping around the node at ${\bm k}_0\in \Sigma$ (see
Fig.~\ref{fig:line-node}). 
Then, the Hamiltonian defines a map, 
$ \bm{k} \in S^{p} \mapsto H(\bm{k})\in \mathcal{Q}$,
from $S^p$ to a classfying space
$\mathcal{Q}$ of matrices subject to symmetries
such as Eqs. (\ref{eq:PHS}), (\ref{eq:TRS}), and (\ref{eq:IS}).
A set of homotopy equivalence class of the map is given by the homotopy
group $\pi_{p} (\mathcal{Q})$. If the node has a nontrivial topological
number of $\pi_{p}(\mathcal{Q})$, we cannot eliminate the node since the
Hamiltonian with the node does not continuously connect to that with a gap.

We may assume here
without loss of generality that the BdG Hamiltonian close to a node
$\bm{k}_0$ is given by 
\begin{align}
 H_{\bm{k}_0} (\bm{p}):= H(\bm{k}_0 + \bm{p}) \simeq \sum_{i= 1}^{p+1} v_i p_i \gamma_i, \label{eq:effBdG}
\end{align}     
where $v_i$ is an expansion coefficient, $|\bm{p} | \ll 1$ and the gamma
matrices $(\gamma_1,\dots, \gamma_{p+1})$ satisfy the Clifford algebra,
$\{ \gamma_i, \gamma_j \} = 2 \delta_{ij}$. $H_{\bm{k}_0} (\bm{p})$
describes a dispersion of $\bm{p}$ near the node, which is determined by
$\epsilon (\bm{k})$ and $\Delta (\bm{k})$ of the underlying BdG
Hamiltonian \cite{comment1}.  
Imposed symmetry of $H_{{\bm k}_0}({\bm p})$ 
depends on whether the node ${\bm k}_0$ is located on a symmetric point
satisfying $\bm{k}_0 = -\bm{k}_0 + \bm{G}$ or not ($\bm{G}$ is a
reciprocal lattice vector). If  $\bm{k}_0 = -\bm{k}_0 + \bm{G}$, all
symmetry operations remain the position of the node unchanged. Thus, the
symmetry operation on $H_{\bm{k}_0}$ is the same as the underlying BdG
Hamiltonian (\ref{eq:BdG2}). On the other hand, if $\bm{k}_0 \neq -
\bm{k}_0 +\bm{G}$, the position of the node changes into its inverse
under C, T, and $\tilde{P}$. Thus, these operations are not the symmetry
of $H_{\bm{k}_0}$. Appropriate symmetries are thus given by the
combination of them such that 
\begin{align} 
   &(C\tilde{P}) H_{\bm{k}_0} (\bm{p}) (C\tilde{P})^{\dagger} = -H_{\bm{k}_0} (\bm{p}), \label{eq:CP} \\
   &(T\tilde{P}) H_{\bm{k}_0} (\bm{p}) (T\tilde{P})^{\dagger} = H_{\bm{k}_0} (\bm{p}). \label{eq:TP}  
\end{align}

On topological stability, nodes at $\bm{k}_0 = -\bm{k}_0 + \bm{G}$ have
been discussed in Refs.~\cite{Zhao:2013v1, Shiozaki:2014}, in which
they took into account PHS and TRS, separately. 
Also, it directly connects to the
Altland-Zirnbauer (AZ) symmetry
classes~\cite{Zirnbauer:1996,Altland:1997} of the bulk electronic
state~\cite{Matsuura:2013,Zhao:2013v2}. 
However, almost all nodes in SCs appear on the Fermi surface, obeying
$\bm{k}_0 \neq -\bm{k}_0
+\bm{G}$. Thus, it is valuable to discuss the stability of nodes with
the symmetries (\ref{eq:CP}) and (\ref{eq:TP}) as a physically
  realistic situation. Hereinafter, we use the combined symmetries to
  classify stable nodes.

{\it PHS, TRS and line node}---
To identify the classifying space of $H_{\bm{k}_0}$, we employ the
Clifford algebra extension method
\cite{Stone:2011,Morimoto:2013,Abramovici:2012}, which enable us to
reduce the problem into an identification of possible Dirac mass terms. 
When returning to
the classification of nodes, the $H_{\bm{k}_0}$ has no mass term;
nevertheless we can apply this method to it by regarding one of the
gamma matrices as the mass term, e.g., $\gamma_{p+1}$, since the base
space $S^p$ is
compactified. According to Eqs.~(\ref{eq:CP}) and (\ref{eq:TP}), we
impose only $C\tilde{P}$ on SCs with IS,
and both
$C\tilde{P}$ and $T\tilde{P}$ on SCs with IS and TRS.
We call the former (latter) systems as P+D (P+DIII) class.
Here the classifying space
depends on either the even-parity ($[C,\tilde{P}]=0$) or the odd-parity
($\{C,\tilde{P}\}=0$).  

Searching for the possible mass terms systematically~\cite{supplement},
we achieve the classifying spaces and the topological numbers for each
class and each codimension as shown in Table~\ref{tab:parity}, in
which we add the topological classification without inversion symmetry
(D and DIII classes) for comparison. We label the classifying spaces as
$C_i$ ($i=0,1$) and $R_j$ ($j=0,1,2,..,7$) according to the conventional
way~\cite{Kitaev:2009,Stone:2011,Wen:2012,Morimoto:2013,Abramovici:2012}. Note
that the higher dimensional homotopy groups in the present case are
calculated by the formula: $\pi_p (C_i) = \pi_0 (C_{i+p})$ and $\pi_p
(R_j) = \pi_0(R_{j+p})$.
In particular, when $p=1$, Table~\ref{tab:parity} shows the stability
of a line node. Hence, a topologically stable line nodes can exist for
the DIII class and the P+DIII class with
even-parity~\cite{Sato:2006,Beri:2010,Schnyder:2011}; this
accounts for the stability of line nodes in noncentrosymmetric SCs such
as {\rm CePt}$_3${\rm Si}~\cite{Izawa:2005,Bonalde:2005} and high-$T_c$
materials~\cite{Hu:1994,Tanaka:1995,Kashiwaya:2000}. On the other hand,
Table~\ref{tab:parity} implies that
a line node in odd-parity SCs
is topologically unstable with or
without TRS.   
The latter statement is one of the main results of the present paper.

{\it Additional symmetry and line node.}---
Now we take into account material dependent symmetries other than IS,
which could stabilize a line node in odd-parity SCs.
In particular, a line node can be invariant under reflection or
spin-rotation, which may give an extra topological obstruction for
opening a gap.

(A) Reflection.--- For simplicity, assuming that the
reflection plane is perpendicular to the $z$-axis,
the reflection symmetry then demands that
\begin{align}
 \tilde{M} H (k_x,k_y,k_z) \tilde{M}^{\dagger} = H (k_x,k_y,-k_z), \label{eq:MS}
\end{align} 
with $\tilde{M} = \diag (M, -\eta^M_C M^{\ast})$. 
The commutation relations between $\tilde{M}$ and $C$, $T$, and
$\tilde{P}$ are defined by $\tilde{M} S = \eta_S^M S \tilde{M}$
($S=C,T,\tilde{P}$), where $\eta^M_S = \pm 1$. 
Without loss of generality, we choose a phase of $M$ so as $M^2 = -1$. 
The reflection can be mirror reflection, which is proper reflection in
the presence of SOC, but the following arguments are
applicable to any kinds of reflection.

We calculate the classifying space by
adding $\tilde{M}$ in the underlying Clifford algebras, where
$\tilde{M}$ satisfies $\{ \gamma_z, \tilde{M}\} = [\gamma_{x,y}
,\tilde{M}]=0$.  
As a result, the topological stability of nodes under the reflection
symmetry is obtained as Table~\ref{tab:spin} (A)~\cite{supplement}, in
which we specify $\eta^M_S$ of $\tilde{M}$ by $M^{\eta_C^M \eta_P^M}$ for the P+D
class and $M^{\eta_C^M \eta_P^M ,\eta_C^M \eta_T^M}$ for the P+DIII
class.
\begin{table}[tbp]
\centering
\caption{Classification of nodes with IS and (A) reflection symmetry or (B) $\pi$-spin-rotational symmetry (SRS). The fourth column of (A) and (B) show types of reflection symmetry class and those of SRS class, respectively. Here the superscripts of $M (U)$ represent the commutation relation with $C \tilde{P}$ and $CT$, i.e., $M^{\eta_C^M \eta_P^M} (U^{\eta_C^U \eta_P^U})$ for the P+D class and $M^{\eta_C^M \eta_P^M, \eta_C^M \eta_T^M} (U^{\eta_C^U \eta_P^U, \eta_C^U \eta_T^U})$ for the P+DIII class. \vspace*{2mm}} \label{tab:spin}

(A) PHS, TRS, IS (odd parity), and reflection symmetry

\begin{tabular}{cccccccc} 
\hline \hline
Class & Symmetry &Parity  & Reflection    &  $\mathcal{Q}$  & $p=0$ & $p=1$ & $p=2$  \\ \hline
P+D  &  \{$C\tilde{P},M$\}& odd   & $M^+$  &  $ R_7 $ & $0$ & $\mathbb{Z}$ & $\mathbb{Z}_2$ \\ 
        &                         &        & $M^-$   & $ R_5$ & $0$ & $0$ & $0$ \\ \hline
P+DIII &\{$C\tilde{P},T\tilde{P},M$\} & odd  &$M^{++}$  & $R_6$& $0$ & $0$ & $\mathbb{Z}$ \\
         &                                 &       & $M^{-+}$  & $R_4$& $\mathbb{Z}$ & $0$ & $0$ \\
        &                                 &    & $M^{+-}$  & $C_1$ & $0$ & $\mathbb{Z}$& $0$ \\
        &                                 &    & $M^{--}$  & $R_5$ & $0$ & $0$ & $0$ \\
 \hline \hline \\
\end{tabular}

(B) PHS, TRS, IS (odd parity), and $\pi$-SRS

 \begin{tabular}{cccccccc} \hline \hline
Class & Symmetry &Parity  & SRS   &  $\mathcal{Q}$  & $p=0$ & $p=1$ & $p=2$ \\ \hline
P+D  &  \{$C\tilde{P},\tilde{U}$\}& odd   & $U^{+}$  &  $ C_0 $ & $\mathbb{Z}$ & $0$ & $\mathbb{Z}$ \\ 
        &                         &        & $U^-$   & $ R_6$ & $0$ & $0$ & $\mathbb{Z}$ \\ \hline
P+DIII &\{$C\tilde{P},T\tilde{P},\tilde{U}$\} & odd  & $U^{++}$  &  $ C_1 $ & $0$ & $\mathbb{Z}$ & $0$ \\
                               &                          &       & $U^{-+}$   & $ R_5 $ & $0$ & $0$ & $0$ \\
                               &                          &       & $U^{+-}$   & $ R_4$ & $\mathbb{Z}$ & $0$ & $0$  \\
         &                          &       & $U^{--}$   & $ R_6$ & $0$ & $0$ & $\mathbb{Z}$  \\
 \hline \hline
\end{tabular}
\end{table}
In the single-band spin-triplet SC with TRS, the symmetry operations
are given by $C= \tau_x K$, $T=i \sigma_y K$ and $\tilde{P} = \tau_z$,
where $\tau_i$ and $\sigma_i$ are the Pauli matrices describing the Nambu
and the spin spaces, respectively, and $K$ represents the complex 
conjugate. Thus, mirror reflection with respect to the $xy$-plane is
labelled as $M^{++}$ ($\bm{d}
\perp \bm{z}$) or $M^{--}$ ($\bm{d} \parallel \bm{z}$);
namely, a line node is unstable as  Blount proved.       
On the other hand, we meet counterexamples of the Blount's argument for
the $M^+$ and the $M^{+-}$ cases~\cite{comment3}; 
The $M^+$ mirror reflection can be realized in a SC without TRS,
whereas $M^{+-}$ can be achieved in a SC with TRS, if they have 
particular normal state and multi-band structures. 
We illustrate them in
the supplement materials \cite{supplement}.      

(B) Spin-rotation.---
For bulk Cooper pairs, the influence of SOC is strongly supressed by the
Fermi energy. Thus, SRS can be an approximately good symmetry. 
For convenience, we consider here $\pi$-SRS such that $[H (\bm{k}),\tilde{U}] = 0$, where $\tilde{U} =
\diag (U, -\eta_C^U U^{\ast})$ and $U^2 =-1$.  
We define the commutation relations between $\tilde{U}$ and $C$, $T$,
and $\tilde{P}$ by $\tilde{U} S = \eta_S^U S \tilde{U}$ and $\eta_S^U =
\pm1$ ($S=C,T,\tilde{P}$). By the Clifford algebra extension method, the
stability of nodes is calculated as shown in
Table~\ref{tab:spin} (B)~\cite{supplement},  where we specify $\eta^U_S$ by
$U^{\eta_C^U \eta_P^U}$ for the P+D class and $U^{\eta_C^U \eta_P^U
,\eta_C^U \eta_T^U}$ for the P+DIII class. From Table~~\ref{tab:spin} (B),
we find a stable line node in the $U^{++}$ class. In the single-band
spin-triplet SC with TRS, the $\pi$-SRS belongs to  $U^{++}$
($\bm{d} \parallel \bm{l}$) or $U^{--}$ ($\bm{d} \perp
\bm{l}$), where $\bm{l}$ is the spin-rotation axis. Thus, the
system may support SRS protected line nodes if $\bm{d}
\parallel \bm{l}$. It is noteworthy that the $U^{++}$ class includes the
polar phase in superfluid $^3$He~\cite{Legget:1975,Blount:1985}.

{\it Surface flat dispersion.}---
\begin{figure}[tbp]
\centering
 \includegraphics[width=8cm]{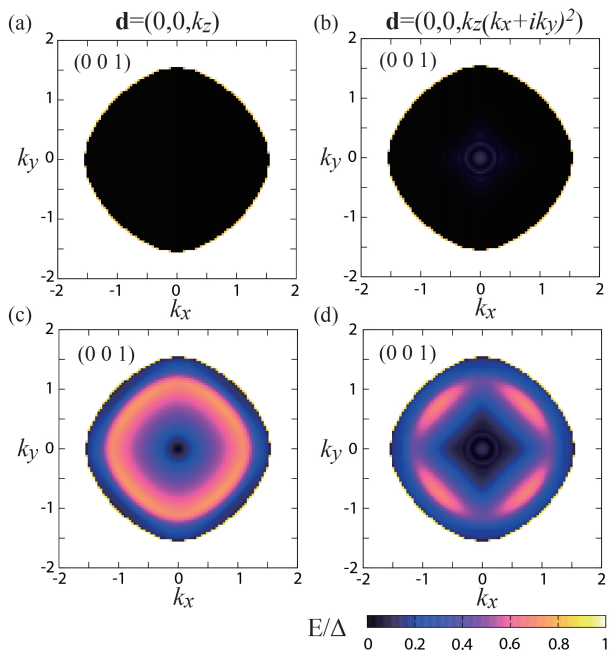}
 \caption{(Color online) Energy spectra at the (001) face of the odd-parity SC with a line node as a function of surface momentum $(k_x,k_y)$.  $\mu = -4t$, $\Delta_0 =0.3t$, and $\lambda = 0.3t$. The distance between two edges is $L =90$. (a),(c) $\bm{d} =\Delta_0(0,0,k_z)$ and  (b),(d) $ \bm{d}=\Delta_0 (0,0,k_z(k_x+ik_y)^2 )$. In (a),(b), we ignore the Rashba SOC, whereas, in (d),(f), we include it in the distance $1\le L  \le 5$ and $85 \le L \le 90$. The color scale shows the energy: the black region represents a zero-energy state. }\label{fig:flat}
\end{figure}
Finally, we discuss implication of our results in experiments.
We first would like to mention that our results do not provide a
strong constraint on the existence of bulk line nodes in odd-parity SCs:
As mentioned in the above,
since SRS could be a good symmetry in the bulk, for odd-parity SCs with
TRS, SRS in
the $U^{++}$ class permits a topological stable bulk line node.
Also, even for those without TRS, the $M^+$ reflection symmetry obtained by
combining mirror reflection with SRS allows to host a stable
bulk line node, as is seen in Table \ref{tab:spin} (A).
Nevertheless, our results do provide a strong implication for
corresponding surface states.
The point is that the surface Rashba SOC, which breaks SRS, can not be
neglected. 
The influence of the surface Rashba SOC is not supressed by the Fermi
energy, and thus the bulk-boundary correspondence does not hold for the SRS
protected line nodes in actual materials.

%
As concrete examples, we numerically calculate the energy spectra for 3D
single-band odd-parity SCs with the gap function of the polar state
\cite{Legget:1975} and the $E_{2u}$ state of UPt$_3$ B-phase \cite{Joynt:2002}. 
The normal state is given by $\epsilon (\bm{k}) = -2 t (\cos k_x +\cos
k_y +\cos k_z) - \mu$, where we assume a spherical Fermi surface,
i.e., $\mu = -4 t$. For the gap function, we consider (a),(c)
$\Delta(\bm{k})=\Delta_0\sin k_z \sigma_x$ for the polar state and
(b),(d) $\Delta(\bm{k})=\Delta_0\sin k_z (\cos k_x +2i \sin k_x \sin k_y
-\cos k_y)\sigma_x$ for the $E_{2u}$ state of UPt$_3$ B-phase. In both cases, a line
node exists on the $k_z=0$ plane. Each line node is protected by (a),(c)
$U^{++}$ in Table \ref{tab:spin} (B) (b),(d) $M^+$ in Table
\ref{tab:spin} (A)~\cite{comment4}. 
The system has the open boundary condition in the
$z$-direction and the periodic boundary condition for the $x$ and $y$
axises. In addition, we take into account the effect of the surface Rashba SOC
as $ \epsilon_{R} (\bm{k}) = \pm  \lambda (\sin k_y \sigma_x - \sin k_x
\sigma_y)$ in a small distance from the open boundary, where we take $+1$
$(-1)$ for the top (bottom) surface. Calculating the surface energy
spectra numerically, we obtain the zero-energy state in the absence of
the surface Rashba SOC, which is described by the black region in (a) and (b) in
Fig.~\ref{fig:flat} \cite{Hara:1986,Sato:2011,Goswami:2014}. However,
once we take the surface Rashba SOC into account, almost all
of zero-energy states disappear for both gap functions (see (c) and (d)
in Fig.~\ref{fig:flat}). This is because the Rashba SOC breaks SRS;
namely, the line node is unstable under the
Rashba SOC and this instability generates a gap in a large region of
the surface state. 
In contrast, the zero-energy surface flat bands in
high-$T_c$ cuprates or noncentrosymmetric SCs are stable under the surface
Rashba SOC since the line nodes are protected by TRS
only~\cite{Sato:2011,Brydon:2011,Schnyder:2012}.  

The instability of the zero-energy state in odd-parity SCs can be tested
by the tunneling spectroscopy as a splitting or broadening of zero-bias
conductance
peak, which gives a clear distinction from the sharp peak in high-$T_c$
materials~\cite{Lofwander:2001,Hu:1994,Tanaka:1995,Kashiwaya:2000,Matsumoto:1995}.

{\it Summary and Discussion.}--- 
We discussed the topological stability of line nodes in odd-parity SCs,
which not only extend the original Blount's theorem but also shows the
counterexamples. Furthermore, the topological arguments give a renovated
meaning of the theorem that a line node associated zero-energy
surface flat band is improbable in odd-parity SCs. 
Our renewed Blount's theorem can be applied
to various unconventional SCs such as UPt$_3$
\cite{Joynt:2002}, UBe$_{13} $\cite{Ott:1984}, UNi$_2$Al$_3$
\cite{Geibel:1991}, Cu$_x$Bi$_2$Se$_3$ \cite{Sasaki:2011}, and so on since
 they are candidates of odd-parity SCs.   
Also, a symmetry protected line node was also proposed for
nonsymmorphic odd-parity SCs \cite{Micklitz:2009}, where the
line node is protected by twofold screw symmetry. 

Whereas we mainly focus on line nodes in odd-parity SCs, our
classification here is also applicable to other nodal structures.  
It is noteworthy here that point nodes
in the $E_{1u}$ state of UPt$_3$ B-phase~\cite{TsuTsumi:2013} and Cu$_x$Bi$_2$Se$_3$~\cite{Yamakage:2012,Yang:2014}
belongs to the $M^{++}$ class in Table~\ref{tab:spin} (A), and
they are topologically stable.

Finally, we would like to mention that our method here works also for Dirac materials such as a
graphene~\cite{Neto:2009}, an organic conductor \cite{Kobayashi:2007},
and so on. For example, if we consider the TRS ($T^2 =1$) and the
inversion symmetry ($P^2 =1$), a combined symmetry is $TP$
($[T,P]=0$). By the same calculation with the superconducting state, we
obtain $\mathcal{Q} =R_0$. The first homotopy group is $\pi_1 (R_0) =
\mathbb{Z}_2$; i.e., the Dirac cone is stable in the 2D systems such as
the graphene. Also, we can predict a stable Dirac cone in a 3D system since $\pi_2 (R_0) = \mathbb{Z}_2$, which will gives a clue of a new topological material.   

S.K. would like to thank A. Yamakage and K. Yada for their helpful comments. This work is supported in part by a Grant-in Aid for Scientific Research from MEXT of Japan, ``Topological Quantum Phenomena,'' Grant No. 22103005 (M.S., Y.T.). S.K. acknowledges support from JPSJ (Grant No. 256466). K.S. is supported by a JSPS Fellowship for Young Scientists.

\clearpage

\onecolumngrid

\renewcommand{\thefigure}{S\arabic{figure}} 

\renewcommand{\thetable}{S\arabic{table}} 

\renewcommand{\thesection}{S\arabic{section}.}

\renewcommand{\theequation}{S.\arabic{equation}}

\setcounter{figure}{0}
\setcounter{table}{0}
\setcounter{equation}{0}

\begin{flushleft} 
{\Large {\bf Supplementary Material}}
\end{flushleft}

\baselineskip24pt

\begin{flushleft} 
{\bf S1. Stability of nodes and topological invariant}
\end{flushleft} 

We discuss the connection between a stability of nodes and a topological invariant. To be concrete, we consider a time-reversal invariant (TRI) superconductor without inversion symmetry. Assuming that a system is three-dimensional and the Fermi surface is spherical, let us first consider a surface node, i.e., $\Delta (\bm{k}) =0$ over the spherical Fermi surface. The Bogoliubov de Gennes (BdG) Hamiltonian is given by
\begin{align}
 H_{\rm BdG} (\bm{k}) = (\bm{k}^2/2m -\mu )\tau_z \otimes 1_{2\times 2},  
\end{align} 
where $m$, $\mu$, and $1_{2 \times 2}$ are a mass of electron, a chemical potential, and a $2 \times 2$ identity matrix, respectively. When we expand $H_{\rm BdG} (\bm{k})$ around a point $\bm{k}_0$ on the Fermi surface, the Hamiltonian close to the node is given by
\begin{align}
   H_{\bm{k}_0} (\bm{p}) := H_{\rm BdG} (\bm{k}_0 + \bm{p}) \simeq \bm{v}_0 \cdot \bm{p} \; \tau_z \otimes 1_{2\times 2}, 
\end{align}
where $|\bm{p}| \ll 1$ and $\bm{v}_0 = \bm{k}_0/m$ ($|\bm{k}_0| = \sqrt{2m\mu}$). To make a superconducting gap on the Fermi surface, it is necessary to find a ``symmetry preserving mass term'' (SPMT) denoted by $\gamma_{\rm M}$ \cite{Ryu:2010s,Chiu:2013s}, which is anticommute with $H_{\bm{k}_0}$. Since the underlying Hamiltonian has particle-hole symmetry (PHS) $C=(\tau_x \otimes 1_{2 \times 2})K$ and time-reversal symmetry (TRS) $T=( 1_{2\times 2} \otimes i\sigma_y)K$ and the node satisfies $\bm{k}_0 \neq -\bm{k}_0$, the SPMT has to satisfy $\{ \gamma_{\rm M}, TC \}=0$. We readily find  the SPMT as $\gamma_{\rm M} = \tau_y \otimes 1_{2\times 2}$. Thus, the surface node is unstable.

 Next, we consider the stability of a line node. The BdG Hamiltonian is given by 
\begin{align}
H_{\rm BdG} (\bm{k}) =  (\bm{k}^2/2m -\mu )\tau_z \otimes 1_{2\times 2} + v_{\Delta} k_z \tau_y \otimes 1_{2 \times 2},
 \end{align}
where $v_{\Delta}$ is an amplitude of the gap function. There is the line node at $k_z =0$ on the Fermi surface. The Hamiltonian close to the nodal point $\bm{k}_0 = (\sqrt{2m\mu}\cos \theta,\sqrt{2m\mu}\sin \theta, 0)$ is given by
\begin{align}
 H_{\bm{k}_0} (\bm{p}) =  \bm{v}_0 \cdot \bm{p} \; \tau_z \otimes 1_{2\times 2} + v_{\Delta} p_z \tau_y \otimes 1_{2 \times 2}.
 \end{align}
In this Hamiltonian, we cannot produce the superconducting gap due to the absence of the SPMT. Therefore, the line node is stable. 

\begin{table*}[tbp]
\centering
 \caption{Bott periodicity of the classifying space for (a) complex case $C_q$ and (b) real case $R_q$. The last columns show the zeroth homotopy group of each classifying space. \vspace*{2mm}} \label{tab:classifying-space}
(a) Complex case

\begin{tabular}{ccc} 
\hline \hline
$q$ mod $2$ & Classifying space $C_q$ & $\pi_0 (C_q)$ \\ \hline
$0$ & $[U(n+m)/U(n) \times U(m)] \times \mathbb{Z}$ & $\mathbb{Z}$ \\
$1$ & $U(n)$ & $0$ \\
 \hline \hline \\
\end{tabular}

(b) Real case

\begin{tabular}{ccc}
\hline \hline
$q$ mod $8$ & Classifying space $R_q$ & $\pi_0 (C_q)$ \\ \hline
$0$ & $[O(n+m)/O(n) \times O(m)] \times \mathbb{Z}$ & $\mathbb{Z}$ \\
$1$ & $O(n)$ & $\mathbb{Z}_2$ \\
$2$ & $O(2n)/U(n)$ & $\mathbb{Z}_2$ \\
$3$ & $U(2n)/Sp(n)$ & $0$ \\
$4$ & $[Sp(n+m)/Sp(n) \times Sp(m)] \times \mathbb{Z}$ & $\mathbb{Z}$ \\
$5$ & $Sp(n)$ & $0$ \\
$6$ & $Sp(n)/U(n)$ & $0$ \\
$7$ & $U(n)/O(n)$ & $0$ \\
 \hline \hline
 \end{tabular}
\end{table*}

As discussed above, the stability of nodes is determined by the existence of the SPMT. In what follows, we show that the stability of nodes relates with a topological invariant. For the sake of completeness, we restricts our attention to the TRI superconductor without inversion symmetry. We note that the original argument of this case is given in Refs.~\cite{Horava:2005s,Zhao:2013s} as the AIII class in the Altland-Zirnbauer symmetry classes.

To see the topological invariant, we assume the Hamiltonian with a sufficiently large matrix dimension and regards the normal dispersion as a ``mass term". Note that we define a mass term to characterize degrees of freedom of $H_{\bm{k}_0}$ based on Refs.~\cite{Kitaev:2009s,Stone:2011s,Wen:2012s,Morimoto:2013s,Abramovici:2012:s}, which is not a real mass term. In the case of the surface node, a $2N \times 2N$ Hamiltonian is given by 
\begin{align}
 H_{\bm{k_p}} (\bm{p}) =  \bm{v}_0 \cdot \bm{p} \; \tau_z \otimes 1_{N\times N}.
\end{align}
The Hamiltonian has a chiral symmetry $\{ H, TC \}  = 0$. Now, we redefine $TC = \tau_x \otimes 1_{N\times N}$ for convenience sake. The general form of the Hamiltonian is given by
\begin{align}
H_{\bm{k_p}}' (\bm{p}) =e^{i \tau_x \otimes A} e^{i \tau_0 \otimes B} (\bm{v}_0 \cdot \bm{p} \; \tau_z \otimes 1_{N\times N}) e^{-i \tau_0 \otimes B} e^{-i \tau_x \otimes A},
\end{align}
where $A$ and $B$ are $N$-by-$N$ Hermitian matrices and $e^{i \tau_x \otimes A} e^{i \tau_0 \times B} \in U(N) \times U(N)$. The Hamiltonian $H'_{\bm{k}_0}$ remains the commutation relation with $TC$ unchanged. Since $[e^{i \tau_0 \times B}, H_{\bm{k_p}}] =0$, the total degrees of freedom of $H_{\bm{k}_0}'$ are $U(N) \times U(N)/U(N) = U(N)$, and which is the classifying space labeled by $C_1$. Since $U(N)$ is the connected space, $\pi_0 (U(N)) =0$. This means that the Hamiltonian with the surface node continuously deforms to that with a full gap. This result is the same as the above argument of the SPMT.  

Secondly, in the case of the line node, a $2N \times 2N$ Hamiltonian is given by
\begin{align}
  H_{\bm{k_p}} (\bm{p}) = \bm{v}_0 \cdot \bm{p} \; \tau_z \otimes C + v_{\Delta} p_z \tau_y \otimes 1_{N \times N},
\end{align} 
where $C$ is a normalized $N \times N$ Hermitian matrices ($C^2 =1_{N \times N}$). The Hamiltonian $H_{\bm{k_p}}$ satisfies $\{H_{\bm{k}_0}, TC\}=0$. Using a $N \times N$ unitary matrix $U_{N \times N} \in U(N)$, $C$ is, in general, given by
\begin{align}
C = U_{N \times N} \diag (1_{n \times n}, -1_{m \times m}) U_{N \times N}^{\dagger}, \ \ m+n =N .
\end{align} 
We readily see that C is invariant under  $\diag (U_{n \times n},U_{m \times m}) \in U(n) \times U(m)$. In addition, we have the freedom of choice about $m \in \mathbb{Z}$. Therefore, the total degrees of freedom of the Hamiltonian is $\bigcup_m [U (n+m)/U(n) \times U(m)]$. In the $N \to \infty$ limit, the classifying space becomes $[U(n+m)/U(n) \times U(m)] \times \mathbb{Z}$, which is formally labeled by $C_0$. Since $\pi_0 (C_0) = \mathbb{Z}$, the Hamiltonian with the line node cannot continuously transform into that with the gap. Thus, the line node is topologically protected as discussed above. Note that the line node is classified by the first homotopy group of $C_1$ because we have the isomorphism: $\pi_0 (C_0) = \pi_1 (C_1)$. We summarize the classifying space $C_q$ and $R_q$ and the zeroth homotopy group of them at Table~\ref{tab:classifying-space}. The higher homotopy groups are determined by the zeroth homotopy group because of the relations: $\pi_p (C_i) = \pi_0(C_{i+p}) = \pi_0 (C_{i+p+2})$ and $\pi_p (R_j) = \pi_0(R_{j+p}) = \pi_0 (R_{j+p+8})$, where the last equality comes from the Bott periodicity~\cite{Karoubi:1978s}.

\begin{flushleft} 
{\bf S2. Clifford algebra extension method}
\end{flushleft} 

In this section, we show the concrete calculation in Tables~\ref{tab:parity} and~\ref{tab:spin}. First, we briefly review a Clifford algebra extension method based on~\cite{Karoubi:1978s,Morimoto:2013s,Abramovici:2012:s}. We define a Clifford algebra $Cl_p$, which has $p$ generators satisfying 
\begin{align}
 \{\gamma_i , \gamma_j\} = 2 \delta_{ij} \ \  (i,j = 1, \cdots p). \label{eq:Cl_complex}
\end{align}
On the other hand, for the real case, a Clifford algebra $Cl_{p,q}$ has $p$ generators satisfying $\gamma_i^2 = -1$ ($i=1,\cdots , p$) and $q$ generators satisfying $\gamma_{p+j}^2 = 1$ ($j=1,\cdots, q$). The generators satisfy the commutation relation such that
\begin{align}
 \{\gamma_i , \gamma_j\} =0 \ \ \text{if } i \neq j. \label{eq:Cl_real}
\end{align} 
For example, $Cl_p$ and $Cl_{p,q}$ are equivalent to the following algebras:
\begin{align}
 Cl_0 = \mathbb{C}, \ \ Cl_1 = \mathbb{C} \oplus \mathbb{C}, 
\end{align}  
and
\begin{align}
 Cl_{0,0} = \mathbb{R}, \ \ Cl_{1,0} = \mathbb{C}, \ \ Cl_{0,1} = \mathbb{R} \oplus \mathbb{R}, \ \ Cl_{2,0} = \mathbb{H}, \ \  Cl_{0,2} = \mathbb{R}(2), 
\end{align}
where $\mathbb{H}$ is a quaternion and $\mathbb{R} (2)$ is a $2 \times 2$ real matrix. We note that ``$=$" represents isomorphism on the algebra. In addition, we have some properties on $Cl_{p,q}$, which is useful to discuss the extension problem, as follows:
\begin{align}
 &Cl_{q,p+2} = Cl_{p,q} \otimes Cl_{0,2}, \label{eq:Clprop1} \\
&Cl_{q+2,p} = Cl_{p,q} \otimes Cl_{2,0}, \label{eq:Clprop2} \\
&Cl_{p+1,q+1} = Cl_{p,q} \otimes Cl_{1,1}, \label{eq:Clprop3} \\
&Cl_{p,q+8} = Cl_{p,q} \otimes Cl_{0,8}, = Cl_{p,q} \otimes \mathbb{R}(16), \label{eq:Clprop4} \\
&Cl_{p+q} = Cl_{p,q} \otimes Cl_{1,0} = Cl_{p,q} \otimes_{\mathbb{R}} \mathbb{C}, \label{eq:Clprop5} \\
&Cl_{p+2} = Cl_{p} \otimes \mathbb{C}(2), \label{eq:Clprop6}
\end{align}
where $\mathbb{R} (16)$ and $\mathbb{C} (2)$ are a $16 \times 16$ real matrix and a $2 \times 2$ complex matrix, respectively.
\begin{table*}[tbp]
\centering
 \caption{Relationship between the symmetry class, the Clifford algebra extension, and the classifying space in the system with inversion symmetry. The first, second, and third columns show the symmetry class, the symmetry constrains, and the parity of gap functions, respectively. The forth and fifth columns show the Clifford algebra extensions and the corresponding classifying spaces. \vspace*{2mm}} \label{tab:ext-parity}
 \begin{tabular}{ccccc} \hline \hline
Class & Symmetry    & Parity  & Extension                     & Classifying space  \\ \hline
D     &   \{$1$\} & N/A & $Cl_{p} \to Cl_{p+1}$    & $C_{p}$ \\
DIII   & \{$TC$\}   &  N/A & $Cl_{p+1} \to Cl_{p+2}$    & $C_{p+1}$ \\ \hline
P+D  & \{$C\tilde{P}$\} & even & $Cl_{0,p+2} \to Cl_{0,p+3}$ & $R_{p+2}$ \\
        &                  & odd & $Cl_{2,p} \to Cl_{2,p+1}$    & $R_{p-2}$  \\ \hline
P+DIII & \{$C\tilde{P},T\tilde{P}$\} & even & $Cl_{0,p+3} \to Cl_{0,p+4}$    & $R_{p+3}$ \\
         &                               & odd &  $Cl_{3,p} \to Cl_{3,p+1}$    & $R_{p-3}$ \\
 \hline \hline
 \end{tabular}
\end{table*}

\begin{table*}[tbp]
\centering
 \caption{Relationship between the symmetry class, the Clifford algebra extension, and the classifying space in the system with inversion symmetry and reflection symmetry. The forth column represents the mirror classes, in which a superscript means a commutation relation with $C$, $T$, and $\tilde{P}$, i.e., $M^{\eta^M_C \eta^M_{P}}$ for P+ D class and $M^{\eta^M_C \eta^M_{P}, \eta^M_C \eta^M_T}$ for P+DIII class, respectively. The fifth and sixth columns show the Clifford algebra extensions and the corresponding classifying spaces for each mirror class. \vspace*{2mm}} \label{tab:ext-mirror}
 \begin{tabular}{cccccc} \hline \hline
Class & Symmetry    & Parity     &Mirror& Extension                     & Classifying space  \\ \hline
P+D  & \{ $C\tilde{P}, \tilde{M}$\} & odd  &$M^+$ &$Cl_{2,p+1} \to Cl_{2,p+2}$ & $R_{p-1}$ \\
        &                                   &       & $M^-$& $Cl_{3,p} \to Cl_{3,p+1}$    & $R_{p-3}$  \\ \hline
P+DIII & \{$C\tilde{P},T\tilde{P},\tilde{M}$\} & odd &  $M^{++}$ & $Cl_{3,p+1} \to Cl_{3,p+2}$    & $R_{p-2}$ \\
         &                               &          & $M^{-+}$ &$Cl_{4,p} \to Cl_{4,p+1}$    & $R_{p-4}$ \\
         &                               &          &  $M^{+-}$ &$Cl_{p+3} \to Cl_{p+4}$    & $C_{p+3}$ \\
         &                               &          & $M^{--}$ & $Cl_{3,p} \to Cl_{3,p+1}$    & $R_{p-3}$ \\
 \hline \hline
 \end{tabular}
\end{table*}
  The Clifford algebra extension method leads the classifying space systematically. The relationship between the Clifford algebra extension and the classifying space is summarized as below:
\begin{align}
 Cl_{p} \to Cl_{p+1} \; &\Leftrightarrow \;C_p, \label{eq:Clext1} \\
Cl_{p,q} \to Cl_{p,q+1} \; &\Leftrightarrow \;R_{q-p}, \label{eq:Clext2} \\
Cl_{p,q} \to Cl_{p+1,q} \; &\Leftrightarrow \;R_{p+2-q}, \label{eq:Clext3}
\end{align}
where the left hand side of Eqs~(\ref{eq:Clext1})-(\ref{eq:Clext3}) represents the Clifford algebra extension and the right hand side of these is the corresponding classifying spaces. The last equation (\ref{eq:Clext3}) is derived from Eq.~(\ref{eq:Clext2}) by using the property (\ref{eq:Clprop1}). Also, we can confirm the Bott periodicity for both the real and complex representations by utilizing the property (\ref{eq:Clprop4}) and (\ref{eq:Clprop6}) since $\mathbb{R} (16) $ and $\mathbb{C} (2)$ do not affect in the extension for each representation.

We show concrete calculations of the Clifford algebra extension method in the system with inversion symmetry, reflection symmetry, and $\pi$-spin-rotation symmetry at Tables~\ref{tab:ext-parity}, ~\ref{tab:ext-mirror}, and~\ref{tab:ext-spin}. For example, in the P+D class with odd parity, i.e., $(C\tilde{P})^2 =-1$, the Hamiltonian of a $(p+1)$-codimensional node is given by
\begin{align} 
H_{\bm{k}_0} (\bm{p}) = \sum_{i=1}^{p+1} k_i \gamma_i, \label{eq:H_p}
\end{align}
which satisfies $\{ H_{\bm{k}_0}, C \tilde{P}\} =0$. In addition, we introduce a ``complex structure" $J$ ($J^2 =-1$), which is anticommutative with $C \tilde{P}$ and is commutative with $H_{\bm{k}_0}$. To see the classifying space of $H_{\bm{k}_0}$, we regard $\gamma_{p+1}$ as a ``mass term". As a result, the Clifford algebras extension is given by $\{\gamma_1 ,\cdots , \gamma_p, C \tilde{P}, J C \tilde{P}\} \to \{\gamma_1 ,\cdots , \gamma_p, \gamma_{p+1},C \tilde{P}, J C \tilde{P}\}$, where $\{\cdots \}$ represents a set of the Clifford algebras satisfying Eq.~(\ref{eq:Cl_real}). This extension means that $Cl_{2,p} \to Cl_{2,p+1}$, so the classifying space is $R_{p-2}$ by Eq.~(\ref{eq:Clext2}). Also, the topological invariant is given by $\pi_0 (R_{p-2}) = \pi_0 (R_{p-2+8}) = \pi_p (R_6)$. 

\begin{table*}[tbp]
\centering
 \caption{Relationship between the symmetry class, the Clifford algebra extension, and the classifying space in the system with inversion symmetry and spin-rotational symmetry (SRS).  The forth column represents the SRS classes, in which a superscript means a commutation relation with $C$, $T$, and $\tilde{P}$, i.e., $U^{\eta^U_C \eta^U_{P}}$ for P+ D class and $M^{\eta^U_C \eta^U_{P}, \eta^U_C \eta^U_T}$ for P+DIII class, respectively. The fifth and sixth columns show the Clifford algebra extensions and the corresponding classifying spaces for each SRS class. \vspace*{2mm}} \label{tab:ext-spin}
 \begin{tabular}{cccccc} \hline \hline
Class & Symmetry    & Parity     &SRS & Extension                     & Classifying space  \\ \hline
P+D  & \{ $C\tilde{P}, \tilde{U}$\} & odd  & $U^+$ &$Cl_{p+2} \to Cl_{p+3}$ & $C_{p+2}$ \\
        &                                   &       & $U^-$& $Cl_{2,p} \to Cl_{2,p+1}$    & $R_{p-2}$  \\ \hline
P+DIII & \{$C\tilde{P},T\tilde{P},\tilde{U}$\} & odd &  $U^{++}$ &$Cl_{p+3} \to Cl_{p+4}$ & $C_{p+3}$ \\
&                               &          &  $U^{-+}$& $Cl_{3,p} \to Cl_{3,p+1}$    & $R_{p-3}$  \\
&                               &          &  $U^{+-}$& $Cl_{4,p} \to Cl_{4,p+1}$    & $R_{p-4}$  \\
         &                               &          &  $U^{--}$& $Cl_{3,p+1} \to Cl_{3,p+2}$    & $R_{p-2}$  \\
 \hline \hline
 \end{tabular}
\end{table*}

When the Hamiltonian (\ref{eq:H_p}) has a reflection symmetry $\tilde{M}$ ($\tilde{M}^2 =-1$) additionally, we need to modify this extension problem. We assume $\{ \gamma_1 ,\tilde{M}\} =[\gamma_{i \neq 1}, \tilde{M}] =[J,\tilde{M}]=0$ so that the reflection affects $k_1$ as $k_1 \to -k_1$, i.e., $k_1$ is momentum transverse to the reflection plane. The Clifford algebra extension depends on whether the reflection symmetry commutes or anticommutes with $C\tilde{P}$. When $[C\tilde{P},\tilde{M}]=0 $ ($M^{+}$ class), the Clifford algebra extension is given by $\{ \gamma_1,\cdots, \gamma_p, C\tilde{P}, J C\tilde{P}, \gamma_1 \tilde{M}\} \to \{ \gamma_1,\cdots,\gamma_p, \gamma_{p+1}, C\tilde{P}, J C\tilde{P}, \gamma_1 \tilde{M}\}$. Hence, the classifying space is $R_{p-1}$. On the other hand, when $\{C\tilde{P},\tilde{M}\}=0$ ($M^-$ class), the Clifford algebra extension is given by   $\{ \gamma_1,\cdots, \gamma_p, C\tilde{P}, J C\tilde{P}, J \gamma_1 \tilde{M}\} \to \{ \gamma_1,\cdots,\gamma_p, \gamma_{p+1}, C\tilde{P}, J C\tilde{P}, J \gamma_1 \tilde{M}\}$. That is, the classifying space is $R_{p-3}$. By repeating the same calculation for each case, we obtain Tables~\ref{tab:ext-parity} and~\ref{tab:ext-mirror}.

Finally, we discuss the Hamiltonian with $\pi$-spin-rotational symmetry $\tilde{U}$, where~$\tilde{U}^2 = -1$ and~$[\gamma_i,\tilde{U}] =[J,\tilde{U}]=0$ ($i=1,2,\cdots, p+1$). Under the $\pi$-spin-rotational symmetry, the Clifford algebra extension is given by $\{ \gamma_1,\cdots,\gamma_p, C\tilde{P}, JC\tilde{P}\} \otimes \{\tilde{U}\} \to \{ \gamma_1,\cdots,\gamma_p, \gamma_{p+1}, C\tilde{P}, JC\tilde{P}\}\otimes \{\tilde{U}\}$   in the $U^+$ class and $\{ \gamma_1,\cdots,\gamma_p, C\tilde{P}, JC\tilde{P}\} \otimes \{J\tilde{U}\} \to \{ \gamma_1,\cdots,\gamma_p, \gamma_{p+1},C\tilde{P}, JC\tilde{P}\}\otimes \{J\tilde{U}\}$ in the $U^-$ class. Here, $\{A\} \otimes \{B\}$ means that $A$ and $B$ are commutative each other. In the former case, the classifying space is $C_{p}$ since $\tilde{U}$ gives the complex structure by Eq.~(\ref{eq:Clprop5}). Whereas, the latter shows the classifying space $R_{p-2}$ since $J\tilde{U}$ just block diagonalizes $H_{\bm{k}_0} $, which has no effect on the classification. In the P+DIII class, we need to include the symmetry $T\tilde{P}$ in the underlying Clifford algebra. For instance, in the $U^{++}$ class, the Clifford algebra extension is given by $\{ \gamma_1 ,\cdots, \gamma_{p} , C\tilde{P},JC\tilde{P},CT\}\otimes \{\tilde{U}\}$ $\to$ $\{ \gamma_1 ,\cdots, \gamma_{p} , \gamma_{p+1},C\tilde{P},JC\tilde{P},CT\}\otimes \{\tilde{U}\}$. Hence, the classifying space is $C_{p+3}$ since $\tilde{U}^2 =-1$. By repeating the same calculation for the other classes, we obtain Table~\ref{tab:ext-spin}.

\begin{flushleft} 
{\bf S3. Examples for the $M^{+}$ and $M^{+-}$ classes}
\end{flushleft} 

As seen in Table II (A), a stable line node is allowed for the $M^{+}$ and $M^{+-}$ classes, which are conflict with the original Blount's argument. In this section, we show that the stability of the line node comes from that of the Fermi surface intersecting with the reflection plane. In what follows, we construct the concrete BdG Hamiltonians belonging to the $M^{+}$ class and the $M^{+-}$ class.

First, we discuss the $M^{+}$ class. The corresponding BdG Hamiltonian is given by
\begin{align}
 H (\bm{k}) = \begin{pmatrix} \epsilon (\bm{k}) -\mu -h \sigma_z & i v_{\Delta} k_z \\ - i v_{\Delta} k_z & - \epsilon (\bm{k}) +\mu +h \sigma_z \end{pmatrix}, \label{eq:H_M+}
\end{align}
where $h$ is a magnetic field of the $z$ direction. The PHS, the inversion symmetry, and the reflection symmetry are given by $C=(\tau_x \otimes 1_{2 \times 2}) K$, $\tilde{P} = \tau_z \otimes 1_{2 \times 2}$, and $\tilde{M}_{xy} = \tau_z \otimes i \sigma_z$, respectively. From the definition, the symmetries satisfy $[C\tilde{P},\tilde{M}] = 0$ and the Hamiltonian (\ref{eq:H_M+}) has the line node at $k_z =0$ on the Fermi surface. On the mirror plane, i.e., $k_z=0$, the Hamiltonian is block diagonalized by $\tilde{M}_{xy}$, whose eigenvalues are given by $\pm i$. Thus, the matrix (\ref{eq:H_M+}) is decomposed into the mirror sector labeled by $H^{(+i)}$ and $H^{(-i)}$ such as 
\begin{align}
H (k_x,k_y,k_z=0) = H^{(+i)} (k_x,k_y,k_z=0)  \oplus H^{(-i)} (k_x,k_y,k_z=0) ,
\end{align}
where
\begin{align}
 H^{(\pm i)} = \pm \begin{pmatrix}\epsilon (\bm{k}) -\mu -h & 0 \\ 0 & -\epsilon (\bm{k}) +\mu -h \end{pmatrix}. \label{eq:h_M+}
\end{align}
Since $H^{(+i)}$ and $H^{(-i)}$ have the same structure, we only consider the $+i$ sector.  The upper and under element of (\ref{eq:h_M+}) represent the Fermi surface of the spin up and the spin down, respectively. When $h > \mu$, the Fermi surface of the spin down component become unstable since $-\epsilon (\bm{k}) +\mu -h =0$ does not have a real solution. In such a situation, there is no perturbation term which produces a superconducting gap. Thus, the line node is stable.

Secondly, we consider the $M^{+-}$ class. The corresponding BdG Hamiltonian is given by
\begin{align}
 H (\bm{k}) = \begin{pmatrix} \epsilon (\bm{k}) -\mu - \lambda \sigma_z \otimes s_x & i v_{\Delta} k_z \\ - i v_{\Delta} k_z & - \epsilon (\bm{k}) +\mu +\lambda \sigma_z \otimes s_x \end{pmatrix}, \label{eq:H_M+-}
\end{align}
where $s_i$ ($i=x,y,z$) is additional degrees of freedom such as an orbital and $\lambda$ is a coupling constant between $\sigma_z$ and $s_x$. The PHS, the TRS, the inversion symmetry, and the reflection symmetry are given by $C=(\tau_x \otimes 1_{4 \times 4}) K$, $T= (1_{2 \times 2} \otimes i \sigma_y \otimes s_z) K$, $\tilde{P} = \tau_z \otimes 1_{4 \times 4}$, and $\tilde{M}_{xy} = \tau_z \otimes i \sigma_z \otimes s_x$, respectively. From the definition, the symmetries satisfy $[C\tilde{P}, \tilde{M}] = \{TC,\tilde{M} \} = 0$ and the Hamiltonian (\ref{eq:H_M+-}) has the line node at $k_z =0$ on the Fermi surface. Since $\tilde{M}^2 =-1$, $H (k_x,k_y,k_z =0)$ is similarly decomposed into $4$-by-$4$ matrices: $H^{(+i)}$ and $H^{(-i)}$. These are given by
\begin{align}
 H^{(\pm i)} = \pm \begin{pmatrix} \epsilon (\bm{k}) -\mu - \lambda & 0 & 0& 0 \\ 0 & \epsilon (\bm{k}) -\mu - \lambda & 0 &0 \\ 0 &0 & - \epsilon (\bm{k}) +\mu +\lambda &0 \\ 0& 0& 0& - \epsilon (\bm{k}) +\mu +\lambda \end{pmatrix}, \label{eq:h_M+-}
\end{align}
where the basis of $H^{(+i)}$ is $(c_{\bm{k},\uparrow 1},c_{\bm{k},\downarrow 2}, c^{\dagger}_{-\bm{k},\uparrow 2},c^{\dagger}_{-\bm{k},\downarrow 1} )$ (the basis of $H^{(-i)}$ consists of the remaining electron states). The subscripts $\uparrow $ ($\downarrow$) and $1$($2$) represent the spin and the additional degrees of freedom, respectively.   When $\lambda > \mu$, the Fermi surface of the electronic states ($c_{-\bm{k},\uparrow 2},c_{-\bm{k},\downarrow 1}$) becomes unstable. In such a situation, Eq.~(\ref{eq:h_M+-}) does not have any perturbation term which produces a gap, so the line node is stable.
 
\begin{flushleft} 
{\bf S4. Zero-energy state and inversion symmetry protected line node}
\end{flushleft} 

In the main paragraph, we show that an inversion symmetry protected line node generates a zero energy flat dispersion at a certain surface numerically. We here prove this statement exactly. In the preceding study, the relation between a line node, which is protected by TRS, and a surface-zero-energy state has been established in Refs.~\cite{Sato:2011s,Schnyder:2011s}. Thus, we here construct a map from a Hamiltonian with inversion symmetry to that without inversion symmetry; namely, we reduce the problem to the topological stability of a line node under the map, which omits the inversion symmetry.  In what follows, we split the main statement into the four statements; (a), (b), (c), and (d) to complete all of the classes in Table I and II. 

First of all, we show that (a) {\it a node is unstable in the D (DIII) class if the node is unstable in the P+D (P+DIII) class}. To see this, we use the following equivalent statements:
\begin{itemize}
\item A topological invariant does not exist.
\item There exists a mass term, which preserves symmetries and is anticommutative with $H_{\bm{k}_0}$. 
\item A node is unstable.
\end{itemize} 
To show (a), we assume that there exists a mass term $\gamma_{M}$ ($[J,\gamma_M] =0$) in the P+D class such that 
\begin{align}
 \{C\tilde{P},\gamma_M\} =\{ H_{\bm{k}_0}, \gamma_M\} =0. \label{eq:mass_P+D}
\end{align}
Also, in the P+DIII class, there exists the mass term satisfying  the following conditions:
 \begin{align}
 \{C\tilde{P},\gamma_M\} =[T\tilde{P},\gamma_M] = \{ H_{\bm{k}_0}, \gamma_M\} =0. \label{eq:mass_P+DIII}
\end{align} 
Alternatively, Eq.~(\ref{eq:mass_P+DIII}) is written by
 \begin{align}
 \{C\tilde{P},\gamma_M\} =\{CT,\gamma_M\} = \{ H_{\bm{k}_0}, \gamma_M\} =0. \label{eq:mass_altP+DIII}
\end{align} 
  By definition, the mass term always makes a gap in the underlying Hamiltonian with or without inversion symmetry. Namely, $\gamma_M$ is the mass term in D (DIII) class as well.

Secondly, we show that (b) {\it when the classifying space becomes the complex class by adding an additional symmetry, a node is stable in the D (DIII) class if the node is stable in the P+D (P+DIII) class}. The proof of this statement consists of three steps: (1) We derive conditions of an additional symmetry $U$ which requires to become the complex class. (2) Both the D and the P+D classes are topologically nontrivial when $p$ is even. Similarly, both the DIII and the P+DIII classes are topologically nontrivial when $p$ is odd. (3) Under the map $f$, which omits the inversion symmetry in the underlying Hamiltonian, a topologically nontrivial Hamiltonian of the P+D (P+DIII) class is mapped into that of the D (DIII) class when $p$ is even (odd). 

In the step (1), the additional symmetry $U$ is defined by
\begin{align}
 \{U, \gamma_i\} = [U,\gamma_{j \neq i}] = [U,J]=0 \ \ (i=1,2,\cdots ,m)
\end{align}
 where $U^2 = \epsilon_U$ and $\epsilon_U = \pm 1$. Then, the condition to become the complex class is directly derived from the Clifford algebra extension method; the results are give by
\begin{itemize}
\item P+D class
\begin{itemize}
\item[(i)] $[C\tilde{P},U]=0$ and $m$ is even, where $m$ satisfies $(-1)^{\frac{m(m+1)}{2}} \epsilon_U = -1$.
\item[(ii)] $\{C\tilde{P},U\}=0$ and $m$ is odd, where $m$ satisfies $(-1)^{\frac{m(m+1)}{2}} \epsilon_U = +1$.
\end{itemize}
\item P+DIII class
 \begin{itemize}
\item[(iii)] $[C\tilde{P},U]=[TC,U]=0$ and $m$ is even, where $m$ satisfies $(-1)^{\frac{m(m+1)}{2}} \epsilon_U = -1$.
\item[(iv)] $\{C\tilde{P},U\}=[TC,U]=0$ and $m$ is even, where $m$ satisfies $(-1)^{\frac{m(m+1)}{2}} \epsilon_U = +1$.
\item[(v)] $[C\tilde{P},U]=\{TC,U\}=0$ and $m$ is odd, where $m$ satisfies $(-1)^{\frac{(m+1)(m+2)}{2}} \epsilon_U \epsilon_{CT} = -1$.
\item[(vi)] $\{C\tilde{P},U\}=\{TC,U\}=0$ and $m$ is odd, where $m$ satisfies $(-1)^{\frac{(m+1)(m+2)}{2}} \epsilon_U \epsilon_{CT} = +1$,
\end{itemize} 
\end{itemize}
The factor $\epsilon_{CT}$ is defined by $(CT)^2 = \epsilon_{CT} = \pm 1$. Note that the mirror symmetry and the $\pi$-spin-rotational symmetry correspond to $m=1$ and $m=0$, respectively. Hence, the $M^{+-}$ class belongs to the case (v), whereas the $U^+$ and the $U^{++}$ classes belong to the cases (i) and (iii) , respectively.

From the calculation of the step (1), the ``complex structure'' $U'$ of the cases (i)-(vi), i.e., $U'$ is commutative with all underlying Clifford algebras, is give by (i),(iii) $U' = \gamma_1 \cdots \gamma_m U$, (ii),(iv) $U' = J \gamma_1 \cdots \gamma_m U$, (v) $U' = CT \gamma_1 \cdots \gamma_m U$, and (vi) $U' = J CT\gamma_1 \cdots \gamma_m U$, respectively. 

Next, to prove the step (2), we relate the P+D (P+DIII) class to the D (DIII) class. This is accomplished by defining a map $f$, which omits the inversion symmetry $\tilde{P}$ in the underlying symmetries. In the D+P class, the map $f$ is given by  
\begin{align}
 f: \{\gamma_1 , \cdots, \gamma_{p+1} ,JC\tilde{P},CP \} \otimes \{ U'\} \to \{\gamma_1 , \cdots, \gamma_{p+1} \} \otimes \{ U'\}. 
\end{align}
 Since the system always belongs to the complex class, the classifying spaces are $C_{p+2}$ in the P+D class and $C_p$ in the D class. Thus, the classifying space is invariant under the map $f$ due to the Bott periodicity. In the same fashion, in the P+DIII class, the map $f$ is defined by
\begin{align}
 f: \{\gamma_1 , \cdots, \gamma_{p+1} ,JC\tilde{P},C\tilde{P},CT \} \otimes \{ U'\} \to \{\gamma_1 , \cdots, \gamma_{p+1},CT \} \otimes \{ U'\}. 
\end{align}
The classifying spaces are $C_{p+3}$ in the P+DIII class and $C_{p+1}$ in the DIII class; i.e., the classifying space remains unchanged under the map $f$. As a result, the step (2) is confirmed.  

Finally, to show the step (3), we construct topologically nontrivial Dirac Hamiltonians of the cases (i)-(vi), which have no SPMT. We describe the Dirac Hamiltonians concretely as follows:
 \begin{itemize}
\item Dirac Hamiltonian of the cases (i) and (ii) \ \ \\
We assume without a loss of generality that $m=0$ and $(C\tilde{P})^2 =-1$. The Dirac Hamiltonians of the case (i) are given by
\begin{align}
 &H_0= k_1 \tau_x, \ \ C\tilde{P} = i \tau_y K, \ \ U = i \tau_x, \notag \\
 &H_2 = H_0 \otimes \sigma_x + k_2 1_{2 \times 2} \otimes \sigma_y + k_3 \tau_x \otimes \sigma_z, \ \ C\tilde{P} = (i \tau_y \otimes 1_{2 \times 2}) K, \ \ U = i \tau_x \otimes 1_{2 \times 2} , \notag \\  
 &\qquad \vdots \notag \\
  &H_{2n} = H_{2n-2} \otimes l_x + k_{2n} 1_{2n\times 2n} \otimes l_y + k_{2n+1} \tau_x \otimes 1_{2n-2 \times 2n-2} \otimes l_z, \notag \\
&\qquad \qquad C\tilde{P} = (i \tau_y \otimes 1_{2n \times 2n}) K, \ \ U = i \tau_x \otimes 1_{2n \times 2n} , \label{D-model1}
\end{align}
where $\tau_i$, $\sigma_i$, and $l_i$ ($i=x,y,z$) are Pauli matrices, respectively.
Similarly, we obtain the case (ii) by replacing $U = i \tau_x \otimes 1_{2n \times 2n}$ with $U = \tau_x \otimes 1_{2n \times 2n}$. 

\item Dirac Hamiltonian of the cases (iii) and (iv) \ \ \\
We assume without a loss of generality that $m=0$, $(C\tilde{P})^2 =-1$ and $(CT)^2 = 1$. The Dirac Hamiltonians of the case (iii) are given by
\begin{align}
 &H_1 = k_1 \tau_x \otimes \sigma_x + k_2 \tau_x \otimes \sigma_z, \notag  \\
& \qquad \qquad C\tilde{P} = (i \tau_y \otimes 1_{2 \times 2}) K, \; U = i \tau_x \otimes 1_{2 \times 2} , \; CT = 1_{2 \times 2} \otimes \sigma_y \notag \\  
&H_3  = H_1 \otimes s_x +k_3 1_{4 \times 4} \otimes s_y + k_4 \tau_x \otimes 1_{2 \times 2} \otimes s_z, \notag \\
&\qquad \qquad  C\tilde{P} = (i \tau_y \otimes 1_{4 \times 4}) K, \; U = i \tau_x \otimes 1_{4 \times 4}, \; CT = 1_{2 \times 2} \otimes \sigma_y \otimes s_x \notag \\
 &\qquad \vdots \notag \\
&H_{2n+1} = H_{2n-1} \otimes l_x + k_{2n+1} 1_{2n+2\times 2n+2} \otimes l_y + k_{2n+2} \tau_x \otimes 1_{2n \times 2n} \otimes l_z, \notag \\
&\qquad \qquad C\tilde{P} = (i \tau_y \otimes 1_{2n+2 \times 2n+2}) K, \; U = i \tau_x \otimes 1_{2n+2 \times 2n+2}, \; CT= 1_{2 \times 2} \otimes \sigma_y \otimes s_x \otimes \cdots \otimes l_x, \label{D-model2}
\end{align}
where $\tau_i$, $\sigma_i$, $s_i$, and $l_i$ ($i=x,y,z$) are Pauli matrices, respectively.
The case (iv) is given by replacing $U = i \tau_x \otimes 1_{2n \times 2n}$ with $U = \tau_x \otimes 1_{2n \times 2n}$..

\item Dirac Hamiltonian of the cases (v) and (vi) \ \ \\
We assume without a loss of generality that $m=1$, $(C\tilde{P})^2 =-1$ and $(CT)^2 = 1$. The Dirac Hamiltonians of the case (vi) are given by
\begin{align}
 &H_1 = k_1 \tau_x \otimes \sigma_x + k_2 \tau_x \otimes \sigma_z,  \notag \\
& \qquad \qquad C\tilde{P} = (i \tau_y \otimes 1_{2 \times 2}) K, \; U = i 1_{2 \times 2} \otimes i\sigma_z , \; CT = 1_{2 \times 2} \otimes \sigma_y \notag \\  
&H_3  = H_1 \otimes s_x +k_3 1_{4 \times 4} \otimes s_y + k_4\tau_x \otimes 1_{2 \times 2} \otimes s_z, \notag \\
&\qquad \qquad C\tilde{P} = (i \tau_y \otimes 1_{4 \times 4}) K, \; U = 1_{2 \times 2} \otimes i \sigma_z \otimes 1_{2 \times 2}, \; CT = 1_{2 \times 2} \otimes \sigma_y \otimes s_x \notag \\
 &\qquad \vdots \notag \\
&H_{2n+1} = H_{2n-1} \otimes l_x + k_{2n+1} 1_{2n+2\times 2n+2} \otimes l_y + k_{2n+2} \tau_x \otimes 1_{2n \times 2n} \otimes l_z, \notag \\
&\qquad \qquad C\tilde{P} = (i \tau_y \otimes 1_{2n+2 \times 2n+2}) K, \; U = 1_{2 \times 2} \otimes i \sigma_z \otimes 1_{2 n-2 \times 2n-2}, \; CT= 1_{2 \times 2} \otimes \sigma_y \otimes s_x \otimes \cdots \otimes l_x, \label{D-model3}
\end{align}
where $\tau_i$, $\sigma_i$, $s_i$, and $l_i$ ($i=x,y,z$) are Pauli matrices, respectively.
The case (v) is given by the same Dirac Hamiltonian with $U = 1_{2 \times 2} \otimes \sigma_z \otimes 1_{2 n-2 \times 2n-2}$. Note that the general forms of Eqs.~(\ref{D-model1}), (\ref{D-model2}), and (\ref{D-model3}) are achieved by acting a unitary operation due to the uniqueness of the Clifford algebras.  
\end{itemize}

As described the above, the higher dimensional Dirac Hamiltonian is inductively derived by the lowest dimensional one. The higher dimensional Dirac Hamiltonian of (i)-(vi) does not have a mass term with or without the inversion symmetry $\tilde{P}$ if there is no mass term in the lowest dimensional one by the property of Pauli matrices. Thus, we focus only on the lowest dimensional one. In the case (i), when we omit the inversion symmetry $\tilde{P}$ in the Hamiltonian $H_0$, the Dirac Hamiltonian and the symmetry become
\begin{align}
 H_0= k_1 \tau_x, \; U = i \tau_x. \label{D-model1-1}
\end{align} 
Obviously, there is no mass term satisfying $\{H_0,\gamma_M\}=[U,\gamma_M]=0$ in Eq.~(\ref{D-model1-1}). Thus, a topologically nontrivial Hamiltonian of the P+D class is mapped to that of the D class. In a similar way, we can verify the absence of the mass term under the map $f$ in the cases (ii)-(vi). As a result, Eqs.~(\ref{D-model1}), (\ref{D-model2}), and (\ref{D-model3}) have no mass term without respect to the inversion symmetry $\tilde{P}$.

Finally, we show that (c) {\it a line node is stable in the D class with mirror symmetry if the line node is stable in the M$^+$ class} and (d) {\it a line node is stable in the DIII class if the line node is stable in the P+DIII with even parity}. In what follows, we attack the statements (c) and (d) individually. 

For the case of (c), the $M^+$ class has the topologically stable line node as shown in Table II (A), whereas the D class with mirror symmetry also has the topological stable line node since $\mathcal{Q} = C_1$ and $\pi_1(C_1) = \mathbb{Z}$. In this case, we can construct the map from a topological nontrivial Hamiltonian of the $M^+$ class to that of the D class with mirror symmetry. To see this, we create the Dirac model of $M^+$ class as below: 
\begin{align}
H_1 = k_1 \tau_z \otimes \sigma_x + k_2 \tau_z \otimes \sigma_z, \; C\tilde{P}= (i \tau_y \otimes 1_{2 \times 2})K, \; \tilde{M} = \tau_z \otimes i \sigma_z. \label{eq:D-model_M+}
 \end{align}   
By Eq.~(\ref{eq:D-model_M+}), the line node is topologically stable regardless of the existence of $C\tilde{P}$. 

Next, for the case of (d), both the P+DIII class with even parity and the DIII class have the topologically stable line node as shown in Table~\ref{tab:parity}. Similarly, we can construct the map from a topologically nontrivial Hamiltonian of the P+DIII class with even parity to that of the DIII class. The Dirac model of this case is  given by
\begin{align}
H_1 = k_1 \tau_z \otimes 1_{2 \times 2} + k_2 \tau_y \otimes \sigma_y, \; C\tilde{P} = (\tau_x \otimes 1_{2 \times 2} ) K , \; CT = \tau_x \otimes i \sigma_y. \label{eq:D-model_even}
\end{align}    
By Eq.~(\ref{eq:D-model_even}), the line node is topologically stable if $C\tilde{P}$ absents.


\begin{thebibliography}{999}
\bibitem{comment1}
We subtract the constant term from $H_{\bm{k}_0}$ and choose an appropriate basis making $H_{\bm{k}_0}$ an irreducible matrix. Also, when dim$H_{\bm{k}_0} = N$, our classification is applicable only in the stable regime: $N > p/2$ for the complex case ($C_i$) and $N > p+1$ for the real case ($R_j$).
\bibitem{comment3}
For the $M^+$ class, the reflection symmetry satisfies $[\tilde{M}, H] = [ \tilde{M}, C ] = 0$ on the reflection plane, namely, $H$ is block diagonalized by an eigenvalue of the reflection symmetry and each sector does not have PHS. The absence of PHS in the sector implies that the stability of the line node is directly determined by that of the Fermi surface intersecting with the reflection plane. In fact, the reflection sector belongs to the topologically nontrivial state of the A class in the AZ classes. Similarly, for the $M^{+-}$ case, the commutation relations are given by $[\tilde{M}, H] = [ \tilde{M}, C ] = \{\tilde{M}, T\} = 0$. Hence, PHS also breaks in the reflection sector, but the TRS remains in each sector, i.e., the reflection sector belongs to the AII class of the AZ classes.
\bibitem{comment4}
For the $E_{2u}$ state of UPt$_3$ B-phase, provided that $\bm{d} \parallel \bm{z}$, the line node is protected by the mirror reflection $\tilde{M} = i \tau_0 \sigma_z$ and the SRS $\tilde{U}_z = i \tau_z \sigma_z$. This is proved as follows. The combined operator $i\tilde{M} \tilde{U}_z = - i\tau_z \sigma_0$ gives a new reflection operator satisfying $(i\tilde{M} \tilde{U}_z)^2 = -1$ and $[C\tilde{P},i\tilde{M}\tilde{U}_z]=0$. Thus, $i\tilde{M}\tilde{U}$ belongs to the $M^+$ class, which has a stable line node.
\end{thebibliography}
\end{document}